\documentstyle[sprocl,epsf]{article}

\def\Journal#1#2#3#4{{#1} {\bf #2}, #3 (#4)}

\def\PLB{{\em Phys.\ Lett.} B}
\def\PRL{\em Phys.\ Rev.\ Lett.}
\def\PRD{{\em Phys.\ Rev.} D}
\def\ZPC{{\em Z.\ Phys.} C}

\begin{document}

\vbox{\rightline{CALT-68-2135} \rightline{UCSD/PTH 97-27} 
  \rightline{hep-ph/9709396} \rightline{}\rightline{} }

\title{SEMILEPTONIC $B$ DECAYS TO EXCITED CHARM MESONS
\footnote{Talk given at Seventh International Symposium On 
Heavy Flavor Physics, July 7--11, 1997, Santa Barbara.  }
}

\author{ZOLTAN LIGETI}

\address{California Institute of Technology, Pasadena, CA 91125 \\ and \\
Department of Physics, University of California San Diego, 
La Jolla, CA 92093~\footnote{Present address.}}

\maketitle\pagestyle{empty}

\vspace{1truecm}\centerline{\bf Abstract}\medskip
\abstracts{Exclusive semileptonic $B$ decays to the lightest excited charmed
mesons are investigated at order $\Lambda_{\rm QCD}/m_Q$ in the heavy quark
effective theory.  At zero recoil, $\Lambda_{\rm QCD}/m_Q$ corrections to the
matrix elements of the weak currents can be written in terms of the leading
Isgur-Wise functions for the corresponding transition and meson mass
splittings.  The differential decay rates are predicted, including
$\Lambda_{\rm QCD}/m_Q$ corrections with some model dependence away from zero
recoil.  Applications to $B$ decay sum rules and factorization are presented.}

\newpage \setcounter{page}{1} \setcounter{footnote}{0} \pagestyle{plain}

\title{SEMILEPTONIC $B$ DECAYS TO EXCITED CHARM MESONS}

\author{ZOLTAN LIGETI}

\address{California Institute of Technology, Pasadena, CA 91125 \\ and \\
Department of Physics, University of California San Diego, 
La Jolla, CA 92093~\footnote{Present address.}}

\maketitle

\abstracts{Exclusive semileptonic $B$ decays to the lightest excited charmed
mesons are investigated at order $\Lambda_{\rm QCD}/m_Q$ in the heavy quark
effective theory.  At zero recoil, $\Lambda_{\rm QCD}/m_Q$ corrections to the
matrix elements of the weak currents can be written in terms of the leading
Isgur-Wise functions for the corresponding transition and meson mass
splittings.  The differential decay rates are predicted, including
$\Lambda_{\rm QCD}/m_Q$ corrections with some model dependence away from zero
recoil.  Applications to $B$ decay sum rules and factorization are presented.}

\section{Introduction}

Heavy quark symmetry~\cite{HQS} implies that in the limit where the heavy quark
mass goes to infinity, matrix elements of the weak currents between a $B$ meson
and an excited charmed meson vanish at zero recoil (where in the rest frame of
the $B$ the final state charmed meson is also at rest).  However, in some cases
at order $\Lambda_{\rm QCD}/m_Q$ these matrix elements are not
zero.~\cite{llsw}  Since most of the phase space for semileptonic $B$ decay to
excited charmed mesons is near zero recoil, $\Lambda_{\rm QCD}/m_Q$ corrections
can be very important.  

In the $m_Q\to\infty$ limit, hadrons containing a single heavy quark, $Q$, are
classified not only by their total spin $J$, but also by the spin of their
light degrees of freedom (i.e., light quarks and gluons), $s_l$.~\cite{IWprl} 
Such hadrons are in degenerate doublets with total spin $J_\pm=s_l\pm\frac12$,
coming from combining the spin of the light degrees of freedom with the spin of
the heavy quark, $s_Q=\frac12$.

The ground state mesons contain light degrees of freedom with spin-parity
$s_l^{\pi_l}=\frac12^-$, giving a doublet containing a spin zero and spin one
meson.  For $Q=c\ (b)$ these are the $D$ and $D^*$ ($B$ and $B^*$) mesons. 
Excited charmed mesons with $s_l^{\pi_l} = \frac32^+$ have been observed. 
These are the $D_1$ and $D_2^*$ with spin one and two, respectively.  In the
nonrelativistic constituent quark model these states correspond to $L=1$
orbital excitations.  Combining the unit of orbital angular momentum with the
spin of the light antiquark leads to states with $s_l^{\pi_l}=\frac12^+$ and
$\frac32^+$.  The $\frac12^+$ doublet, $(D_0^*,D_1^*)$, has not been observed,
presumably because these states are much broader than those with
$s_l^{\pi_l}=\frac32^+$.

The hadron masses give important information on some HQET matrix elements.  The
mass formula for a spin symmetry doublet of hadrons $H_\pm$ with total spin
$J_\pm=s_l\pm\frac12$ is
\begin{equation}\label{mass}
m_{H_\pm} = m_Q + \bar\Lambda^H - {\lambda_1^H \over 2 m_Q} 
  \pm {n_\mp\, \lambda_2^H \over 2m_Q} + \ldots \,,
\end{equation}
where $n_\pm = 2J_\pm+1$ is the number of spin states in the hadron $H_\pm$. 
$\bar\Lambda$ is the energy of the light degrees of freedom in the
$m_Q\to\infty$ limit, $\lambda_1$ determines the heavy quark kinetic energy,
and $\lambda_2$ determines the chromomagnetic energy.  These parameters depend
on the particular spin symmetry doublet to which $H_\pm$ belong.  We reserve
the notation $\bar\Lambda$, $\lambda_1$, $\lambda_2$ for the ground state
multiplet and use $\bar\Lambda'$, $\lambda_1'$, $\lambda_2'$ and
$\bar\Lambda^*$, $\lambda_1^*$, $\lambda_2^*$ for the excited
$s_l^{\pi_l} = \frac32^+$ and $\frac12^+$ doublets, respectively.  The
excitation energy $\bar\Lambda'-\bar\Lambda$ plays a very important role,
\begin{equation}\label{HQET_diff}
\bar\Lambda' - \bar\Lambda = {m_b\,(\overline{m}_B'-\overline{m}_B)
  - m_c\,(\overline{m}_D'-\overline{m}_D) \over m_b-m_c} 
  + {\cal O}\bigg({\Lambda_{\rm QCD}^3\over m_Q^2}\bigg)
  \simeq 0.39\, {\rm GeV} \,,
\end{equation}
where $\overline{m}_H=(n_+m_{H_+}+n_-m_{H_-})/(n_++n_-)$ are the spin average
masses.  This value of $\bar\Lambda'-\bar\Lambda$ has only a small sensitivity
to $m_b$ and $m_c$ (we used $m_b=4.8\,$GeV and $m_c=1.4\,$GeV), but it has
considerable uncertainty because the experimental error on $\overline{m}_B'$ is
large.  (We identified the $B^{(*)}\pi$ resonances observed at LEP with the
bottom $s_l^{\pi_l}=\frac32^+$ meson doublet,
$\overline{m}_B'=5.73\,$GeV.~\cite{talks})

\section{The $B\to D_1e\bar\nu_e$ and $B\to D_2^*e\bar\nu_e$ decays}

The matrix elements of the vector and axial-vector currents 
($V^\mu=\bar c\,\gamma^\mu\,b$ and $A^\mu=\bar c\,\gamma^\mu\gamma_5\,b$) 
between $B$ mesons and $D_1$ or $D_2^*$ mesons are parameterized as
\begin{eqnarray}\label{formf1}
{\langle D_1(v',\epsilon)|\, V^\mu\, |B(v)\rangle \over \sqrt{m_{D_1}\,m_B}}
  &=& f_{V_1}\, \epsilon^{*\mu} 
  + (f_{V_2} v^\mu + f_{V_3} v'^\mu)\, (\epsilon^*\cdot v) \,, \nonumber\\*
{\langle D_1(v',\epsilon)|\, A^\mu\, |B(v)\rangle \over \sqrt{m_{D_1}\,m_B}}
  &=& i\, f_A\, \varepsilon^{\mu\alpha\beta\gamma} 
  \epsilon^*_\alpha v_\beta v'_\gamma \,, \nonumber\\
{\langle D^*_2(v',\epsilon)|\, A^\mu\, |B(v)\rangle \over\sqrt{m_{D_2^*}\,m_B}}
  &=& k_{A_1}\, \epsilon^{*\mu\alpha} v_\alpha 
  + (k_{A_2} v^\mu + k_{A_3} v'^\mu)\,
  \epsilon^*_{\alpha\beta}\, v^\alpha v^\beta \,, \nonumber\\*
{\langle D^*_2(v',\epsilon)|\, V^\mu\, |B(v)\rangle \over\sqrt{m_{D_2^*}\,m_B}}
  &=& i\,k_V\, \varepsilon^{\mu\alpha\beta\gamma} 
  \epsilon^*_{\alpha\sigma} v^\sigma v_\beta v'_\gamma \,, 
\end{eqnarray}
where the form factors $f_i$ and $k_i$ are dimensionless functions of $w=v\cdot
v'$.  At zero recoil ($v=v'$) only the $f_{V_1}$ form factor can contribute,
since $v'$ dotted into the polarization ($\epsilon^{*\mu}$ or
$\epsilon^{*\mu\alpha}$) vanishes.

To write the differential decay rates in terms of the form factors in
Eq.~\ref{formf1}, we define $\theta$ as the angle between the charged lepton
and the charmed meson in the rest frame of the virtual $W$ boson, i.e., in the
center of momentum frame of the lepton pair.  The different helicities of the
$D_1$ or $D_2^*$ yield distinct distributions in $\theta$, which will probably
be measurable.  These helicity amplitudes are affected by the $1/m_Q$
corrections in different ways, so measuring them will be interesting.  In terms
of $w=v\cdot v'$ and $\cos\theta$, the differential decay rates are 
\begin{eqnarray}
{{\rm d}\Gamma_{D_1}\over\Gamma_0} &=& 
  3 r_1^3\, \sqrt{w^2-1}\, \bigg\{ \sin^2\theta\,
  \Big[ (w-r_1) f_{V_1}+(w^2-1) (f_{V_3}+r_1 f_{V_2}) \Big]^2 \nonumber\\*
&&{} + (1-2r_1w+r_1^2) \Big[ (1+\cos^2\theta)\, [f_{V_1}^2 + (w^2-1) f_A^2] 
  \nonumber\\*
&& \phantom{+(1-2r_1w+r_1^2)}
  - 4\cos\theta\, \sqrt{w^2-1}\, f_{V_1}\, f_A \Big] \bigg\}\,
  {\rm d}w\,{\rm d}\!\cos\theta \,,\label{rate1}\\
{{\rm d}\Gamma_{D_2^*}\over\Gamma_0} &=& 
   2 r_2^3\, (w^2-1)^{3/2}\, \bigg\{ \sin^2\theta\,
  \Big[ (w-r_2) k_{A_1}+(w^2-1) (k_{A_3}+r_2 k_{A_2}) \Big]^2 \nonumber\\*
&&{} + \frac34\, (1-2r_2w+r_2^2) \Big[ (1+\cos^2\theta)\, 
  [k_{A_1}^2 + (w^2-1) k_V^2] \nonumber\\*
&& \phantom{+\frac34\,(1-2r_1w+r_1^2)}
  - 4\cos\theta\, \sqrt{w^2-1}\, k_{A_1}\, k_V \Big] \bigg\}\,
  {\rm d}w\,{\rm d}\!\cos\theta \,, \label{rate2}
\end{eqnarray}
where $\Gamma_0={G_F^2\,|V_{cb}|^2\,m_B^5 /(192\pi^3)}$, $r_1=m_{D_1}/m_B$ and
$r_2=m_{D_2^*}/m_B$.  The semileptonic $B$ decay rate into any $J\neq1$ state
involves an extra factor of $w^2-1$.  The $\sin^2\theta$ term is the helicity
zero rate, while the $1+\cos^2\theta$ and $\cos\theta$ terms determine the
helicity $\lambda=\pm1$ rates.  Since the weak current is $V-A$ in the standard
model, $B$ mesons can only decay to the helicity $|\lambda|=0,1$ components of
any excited charmed mesons.  The decay rate for $|\lambda|=1$ vanishes at
maximal recoil, $w_{\rm max}=(1+r^2)/(2r)$, as implied by the $1-2rw+r^2$
factors above ($r=r_1$ or $r_2$).  The differential decay rates in terms of the
electron energy follow from Eqs.~\ref{rate1} and \ref{rate2} using 
$y \equiv 2E_e/m_B = 1 -rw -r\sqrt{w^2-1}\,\cos\theta$.

The form factors $f_i$ and $k_i$ can be parameterized by a set of Isgur-Wise
functions at each order in $\Lambda_{\rm QCD}/m_Q$.  In the $m_Q\to\infty$
limit, $f_i$ and $k_i$ are given in terms of a single function
$\tau(w)$.~\cite{IWsr}  Only $f_{V_1}$ can contribute to the matrix elements in
Eq.~\ref{formf1} at zero recoil.  Heavy quark symmetry does not fix $\tau(1)$,
since $f_{V_1}(1)=0$ in the infinite mass limit independent of the value of
$\tau(1)$.  

At order $1/m_Q$ several new Isgur-Wise functions occur.  Corrections from
matching QCD currents onto HQET currents introduce dependence on
$\bar\Lambda'$, $\bar\Lambda$, and two new functions, $\tau_{1,2}(w)$.  The
order $1/m_c$ chromomagnetic correction to the Lagrangian introduces three new
functions, $\eta_{1,2,3}(w)$, while the charm quark kinetic energy term
introduces dependence of the form factors on one more function, $\eta_{\rm
ke}(w)$.  The functions parameterizing $1/m_b$ corrections to the Lagrangian
occur in a single linear combination, $\eta_b(w)$.  Denoting
$\varepsilon_{c,b}=1/2m_{c,b}$, the $B\to D_1e\bar\nu_e$ form factors including
order $1/m_{c,b}$ terms are~\cite{llsw}
\begin{eqnarray}\label{expf}
\sqrt6\, f_A &=& - (w+1)\tau 
  - \varepsilon_b \{ (w-1) [(\bar\Lambda'+\bar\Lambda)\tau 
  - (2w+1)\tau_1-\tau_2] + (w+1)\eta_b \} \nonumber\\*
&& - \varepsilon_c [ 4(w\bar\Lambda'-\bar\Lambda)\tau - 3(w-1) (\tau_1-\tau_2) 
  + (w+1) (\eta_{\rm ke}-2\eta_1-3\eta_3) ] \,,\nonumber\\*
\sqrt6\, f_{V_1} &=&  (1-w^2)\tau 
  - \varepsilon_b (w^2-1) [(\bar\Lambda'+\bar\Lambda)\tau 
  - (2w+1)\tau_1-\tau_2 + \eta_b] \nonumber\\*
&& - \varepsilon_c [ 4(w+1)(w\bar\Lambda'-\bar\Lambda)\tau
  - (w^2-1)(3\tau_1-3\tau_2-\eta_{\rm ke}+2\eta_1+3\eta_3) ] \,, \nonumber\\
\sqrt6\, f_{V_2} &=& -3\tau - 3\varepsilon_b [(\bar\Lambda'+\bar\Lambda)\tau 
  - (2w+1)\tau_1-\tau_2 + \eta_b] \nonumber\\* 
&& - \varepsilon_c [ (4w-1)\tau_1+5\tau_2 +3\eta_{\rm ke} +10\eta_1 
  + 4(w-1)\eta_2-5\eta_3 ] \,, \nonumber\\*
\sqrt6\, f_{V_3} &=&  (w-2)\tau 
  + \varepsilon_b \{ (2+w) [(\bar\Lambda'+\bar\Lambda)\tau 
  - (2w+1)\tau_1-\tau_2] - (2-w)\eta_b \} \nonumber\\*
&& + \varepsilon_c [ 4(w\bar\Lambda'-\bar\Lambda)\tau + 
  (2+w)\tau_1 + (2+3w)\tau_2 \nonumber\\*
&& \phantom{\varepsilon_c [}
  + (w-2)\eta_{\rm ke} - 2(6+w)\eta_1 - 4(w-1)\eta_2 - (3w-2)\eta_3 ] \,. 
\end{eqnarray}
The analogous formulae for $B\to D_2^*e\bar\nu_e$ are
\begin{eqnarray}\label{expk}
k_V &=& - \tau - \varepsilon_b [(\bar\Lambda'+\bar\Lambda)\tau 
  - (2w+1)\tau_1-\tau_2 + \eta_b] \nonumber\\*
&& - \varepsilon_c (\tau_1-\tau_2+\eta_{\rm ke}-2\eta_1+\eta_3) \,,\nonumber\\*
k_{A_1} &=& - (1+w)\tau - \varepsilon_b \{ (w-1)
  [(\bar\Lambda'+\bar\Lambda)\tau - (2w+1)\tau_1-\tau_2] + (1+w)\eta_b \} 
  \nonumber\\*
&& - \varepsilon_c [ (w-1)(\tau_1-\tau_2)
  + (w+1)(\eta_{\rm ke}-2\eta_1+\eta_3) ] \,, \nonumber\\
k_{A_2} &=& - 2\varepsilon_c (\tau_1+\eta_2) \,, \nonumber\\*
k_{A_3} &=& \tau + \varepsilon_b [(\bar\Lambda'+\bar\Lambda)\tau 
  - (2w+1)\tau_1-\tau_2 + \eta_b] \nonumber\\*
&& - \varepsilon_c (\tau_1+\tau_2-\eta_{\rm ke}+2\eta_1-2\eta_2-\eta_3) \,. 
\end{eqnarray}

At order $1/m_Q$ the form factor $f_{V_1}(1)$ is no longer zero.  At this order
it can be written in terms of $\bar\Lambda'-\bar\Lambda$ and the Isgur-Wise
function $\tau(w)$ evaluated at zero recoil. Explicitly,~\cite{llsw}
\begin{equation}\label{fV1}
\sqrt{6}\, f_{V_{1}}(1) = 
  - {4\over m_c}\, (\bar\Lambda'-\bar\Lambda)\, \tau (1)\,.
\end{equation}
The factor of four in the numerator of Eq.~\ref{fV1} makes this quite a large
correction.  Furthermore, its importance is enhanced over other $\Lambda_{\rm
QCD}/m_Q$ corrections since most of the phase space is near zero recoil.  (For
a flavor diagonal current a relation similar to Eq.~\ref{fV1} was previously 
obtained by Voloshin.~\cite{Volo})

The allowed kinematic range for $B\to D_1e\bar\nu_e$ ($B\to D_2^*e\bar\nu_e$)
decay is $1<w<1.32$ ($1<w<1.31$).  Since these ranges are fairly small, and
there are some constraints on the $1/m_Q$ corrections at zero recoil, it is
useful to consider the decay rates given in Eqs.~\ref{rate1} and \ref{rate2}
expanded in powers of $w-1$.  The general structure of the expansion of ${\rm
d}\Gamma/{\rm d}w$ is elucidated schematically below,
\begin{eqnarray}\label{schematic}
{{\rm d}\Gamma_{D_1}^{(\lambda=0)}\over {\rm d}w} &\sim&
  \sqrt{w^2-1}\, \Big[ ( 0
  + 0\,\varepsilon + \varepsilon^2 + \varepsilon^3 + \ldots )
  + (w-1)\, ( 0 + \varepsilon + \varepsilon^2 + \ldots ) \nonumber\\*
&& \phantom{\sqrt{w^2-1}}
  + (w-1)^2\, ( 1 + \varepsilon + \ldots ) + \ldots \Big] \,, 
  \nonumber\\*
{{\rm d}\Gamma_{D_1}^{(|\lambda|=1)}\over {\rm d}w} &\sim&
  \sqrt{w^2-1}\, \Big[ ( 0 
  + 0\,\varepsilon + \varepsilon^2 + \varepsilon^3 + \ldots ) 
  + (w-1)\, ( 1 + \varepsilon + \ldots )  \nonumber\\*
&& \phantom{\sqrt{w^2-1}}
  + (w-1)^2\, ( 1 + \varepsilon + \ldots ) + \ldots \Big] \,, \nonumber\\*
{{\rm d}\Gamma_{D_2^*}^{(|\lambda|=0,1)}\over {\rm d}w} &\sim&
  (w^2-1)^{3/2} \Big[ ( 1 + \varepsilon + \ldots ) 
  + (w-1) ( 1 + \varepsilon + \ldots ) + \ldots \Big] \,. 
\end{eqnarray}
Here $\varepsilon^n$ denotes a term of order $(\Lambda_{\rm QCD}/m_Q)^n$.  The
zeros in Eq.~\ref{schematic} are consequences of heavy quark symmetry, as the
leading contribution to the matrix elements of the weak currents at zero recoil
is of order $\Lambda_{\rm QCD}/m_Q$.  Thus, the $D_1$ decay rate at $w=1$
starts out at order $\Lambda_{\rm QCD}^2/m_Q^2$.  Similarly, the vanishing of
$f_{V_1}(1)$ in the $m_Q\to\infty$ limit implies that at order $w-1$ the
$D_1^{(\lambda=0)}$ rate starts out at order $\Lambda_{\rm QCD}/m_Q$.  The
$D_2^*$ decay rate is suppressed by an additional power of $w^2-1$, so there is
no further restriction on its structure.

\section{Applications}

\subsection{Predictions for $B\to(D_1,D_2^*)e\bar\nu_e$ decays}

Predictions for various quantities can be made assuming a linear dependence of
the Isgur-Wise function, $\tau(w)=\tau(1)[1+\hat\tau'\,(w-1)]$.  Lagrangian
corrections from the chromomagnetic term are expected to be small, so
$\eta_{1,2,3}$ are neglected in Eqs.~\ref{expf} and \ref{expk}.  In these
equations $\eta_{\rm ke}$ and $\eta_b$ are also set to zero, since the
contributions from the kinetic energy operator can be absorbed into $\tau$. 
Then the unknown parameters entering our predictions are $\hat\tau'$, $\tau_1$
and $\tau_2$.  Approximations B$_1$ and B$_2$ correspond to two choices of
$\tau_1$ and $\tau_2$ ($\tau_1=\tau_2=0$ for B$_1$; $\tau_1=\bar\Lambda\,\tau$
and $\tau_2=-\bar\Lambda'\,\tau$ for B$_2$).  The difference between these
approximations gives a rough estimate of the uncertainty due to unknown $1/m_Q$
corrections.

Recently the ALEPH~\cite{ALEPH} and CLEO~\cite{CLEO} Collaborations measured,
with some assumptions, the $B\to D_1e\bar\nu_e$ branching ratio.  The 
average of their results is
\begin{equation}\label{data}
  {\cal B}(B\to D_1\,e\,\bar\nu_e) = (6.0\pm1.1) \times 10^{-3} \,.  
\end{equation}
The $B\to D_2^*\,e\,\bar\nu_e$ branching ratio has not yet been measured; 
CLEO and ALEPH set upper bounds on ${\cal B}(B\to~D_2^*e\bar\nu_e)$ at the 
1\% level.

Table~\ref{tab:narrows} presents $R=\Gamma_{D_2^*}/\Gamma_{D_1}$ and the value
of $\tau(1)$ extracted from Eq.~\ref{data}, both in the infinite mass limit
where $\varepsilon_c=\varepsilon_b=0$ and for the physical values of
$\varepsilon_c=1/(2.8\,{\rm GeV})$ and $\varepsilon_b=1/(9.6\,{\rm GeV})$.  In
this table $\hat\tau'(1)=-1.5$ is used, motivated by model
calculations.~\cite{ISGW,Cola,VeOl,More}  The $1/m_Q$ corrections enhance the
$D_1$ rate significantly, while they affect the $D_2^*$ rate to a much smaller
extent.  Thus, the values of $\tau(1)$ and $R$ are substantially reduced from
their values in the infinite mass limit.  $R$ is fairly insensitive to
$\hat\tau'$ and $\tau_2$, but depends strongly on $\tau_1$.  $\tau(1)$ is
insensitive to $\tau_2$, but $\hat\tau'$ and $\tau_1$ affect it at the 20\%
level.  For a more detailed discussion and other predictions, see Ref.~[2].
\begin{table}[t]
\caption[2]{Predictions for the ratio of $B\to D_1e\bar\nu_e$ and 
$B\to D_2^*e\bar\nu_e$ decay rates, and the extracted value of $\tau(1)$.  
These results correspond to $\hat\tau'=\tau'(1)/\tau(1)=-1.5$.} 
  \begin{center}
  \begin{tabular}{|c|cc|} 
\hline
Approximation  &  ~~$R=\Gamma_{D_2^*}/\Gamma_{D_1}$~  &  
  ~$\tau(1)\, \bigg[\displaystyle {6.0\times10^{-3} \over 
    {\cal B}(B\to D_1e\bar\nu_e)} \bigg]^{1/2}$~~  \\ \hline 
$m_Q\to\infty$  &  $1.65$  &  $1.24$  \\
Finite~$m_Q\, \bigg\{ \matrix{{\rm B}_1 \cr {\rm B}_2}$  
  &  $\matrix{0.52 \cr 0.67}$  &  $\matrix{0.71 \cr 0.75}$  \\ \hline
  \end{tabular} 
  \end{center}
\label{tab:narrows}
\end{table}

With more experimental information on the differential decay rates for $B\to
D_1e\bar\nu_e$ and $B\to D_2^*e\bar\nu_e$ it should be possible to determine
from experiment the values of $\hat\tau'$ and $\tau_1$.  This would remove much
of the uncertainty in the predictions presented in Table~\ref{tab:narrows}.

\subsection{Factorization}

Factorization should be a good approximation for $B$ decay to charmed mesons
and a charged pion.  Neglecting the pion mass, the two-body decay rate,
$\Gamma_\pi$, is related to the differential decay rate ${\rm d}\Gamma_{\rm
sl}/{\rm d}w$ at maximal recoil for the semileptonic decay with the $\pi$
replaced by the $e\bar\nu_e$ pair.  This relation is independent of the
identity of the charmed meson in the final state,
\begin{equation}\label{factor}
\Gamma_\pi = {3\pi^2\, |V_{ud}|^2\, C^2\, f_\pi^2 \over m_B^2\, r} \times
  \left( {{\rm d} \Gamma_{\rm sl}\over {\rm d}w} \right)_{w_{\rm max}} .
\end{equation}
Here $r$ is the mass of the charmed meson divided by $m_B$, $w_{\rm
max}=(1+r^2)/(2r)$, and $f_\pi\simeq132\,$MeV is the pion decay constant.  $C$
is a combination of Wilson coefficients of four-quark operators, and
numerically $C\,|V_{ud}|\simeq1$.  These nonleptonic decay rates can therefore
be predicted from ${\rm d}\Gamma_{\rm sl}/{\rm d}w$ at maximal recoil.  In the
absence of a measurement of the differential decay rates, we can use our
results for the shape of ${\rm d}\Gamma_{\rm sl}/{\rm d}w$ to predict the $B\to
D_1\pi$ and $B\to D_2^*\pi$ rates.  

At present there are only crude measurements of the ${\cal B}(B\to D_1\pi)$ and
${\cal B}(B\to D_2^*\pi)$ branching ratios.  Assuming ${\cal B}(D_1(2420)^0\to
D^{*+}\pi^-)=2/3$ and ${\cal B}(D_2^*(2460)^0\to D^{*+}\pi^-)=0.2$, the
measured rates are \cite{CLEOfact}
\begin{eqnarray}\label{factordata}
{\cal B}(B^-\to D_1(2420)^0\pi^-) &=& (1.17\pm0.29) \times 10^{-3} \,, 
  \nonumber\\*
{\cal B}(B^-\to D_2^*(2460)^0\pi^-) &=& (2.1\pm0.9) \times 10^{-3} \,.
\end{eqnarray}

Our prediction for ${\cal B}(B\to D_1\pi)/{\cal B}(B\to D_1e\bar\nu_e)$ varies
between 0.13 and 0.31 depending on $\hat\tau'$, fairly independent of
$\tau_{1,2}$.  Assuming that factorization works at the 10\% level, a precise
measurement of the ${\cal B}(B\to D_1\pi)$ rate may provide a determination of
$\hat\tau'$.  The present experimental data, ${\cal B}(B\to D_1\pi)/{\cal
B}(B\to D_1e\bar\nu_e)\simeq0.2$, does in fact support $\hat\tau'\sim-1.5$,
which we took as the ``central value", motivated by model
calculations.~\cite{llsw}

The prediction for ${\cal B}(B\to D_2^*\pi)/{\cal B}(B\to D_1\pi)$, on the
other hand, only weakly depends on $\hat\tau'$, but it is more sensitive to
$\tau_2$ and especially $\tau_1$.  Varying $\tau_{1,2}$, we can accommodate
almost any value of ${\cal B}(B\to D_2^*\pi)/{\cal B}(B\to D_1\pi)$ between 0
and 1.5.  In the future, experimental data on this ratio and $R$ may lead to a
determination of $\hat\tau_1$.

\subsection{$B\to(D_0^*,D_1^*)e\bar\nu_e$ decays}

Similar results to those presented in Sec.~2 also hold for decays to the
$s_l^{\pi_l}=\frac12^+$ doublet, $B\to(D_0^*,D_1^*)e\bar\nu_e$.~\cite{llsw} 
The zero recoil matrix elements are determined by the excitation energy,
$\bar\Lambda^*-\bar\Lambda\simeq0.35\,$GeV, and the $m_Q\to\infty$ Isgur-Wise
function for these transitions, $\zeta(w)$, at zero recoil.  Numerically, the
$D_0^*$ rate is enhanced similar to the $D_1$ rate, whereas $1/m_Q$ corrections
to the $D_1^*$ rate enter near $w=1$ proportional to the anomalously small
combination $\varepsilon_c-3\varepsilon_b$.

To obtain even a crude absolute prediction for the $B\to D_1^*,D_0^*$ rates, a
relation between the $s_l^{\pi_l}=\frac12^+$ and $\frac32^+$ Isgur-Wise
functions is needed.  In any nonrelativistic constituent quark model with
spin-orbit independent potential, $\zeta$ and $\tau$ are related by
\cite{VeOl,IWsr}
\begin{equation}\label{ztau}
  \zeta(w) = {w+1\over\sqrt3}\, \tau(w) \,,
\end{equation}
since both the $\frac12^+$ and $\frac32^+$ doublets correspond to $L=1$ orbital
excitations.  

This relation implies that the six lightest charmed mesons contribute about
8.2\% of the $B$ decay rate ($D$ and $D^*$ is about $6.6\%$ of the total $B$
decay rate~\cite{PDG}).  Therefore, semileptonic decays into higher excited
states and non-resonant multi-body channels should be at least 2\% of the $B$
decay rate, and possibly around 3\% if the semileptonic $B$ branching ratio is
closer to the LEP result of about 11.5\%.  Such a sizable contribution from
higher mass charmed mesons and non-resonant modes would soften the lepton
spectrum, and may make the agreement with data on the inclusive lepton spectrum
worse.  If $\zeta$ were enhanced by a factor of two compared to Eq.~\ref{ztau},
then semileptonic $B$ decay rate to the six lightest charmed mesons could add
up to close to $10\%$.  However, model calculations \cite{More} seem to obtain
a suppression rather than an enhancement of $\zeta$ compared to Eq.~\ref{ztau}.
Thus, taking the measurements for the $B\to D$, $D^*$, and $D_1$ semileptonic
branching ratios on face value, a decomposition of the semileptonic rate as a
sum of exclusive channels seems problematic.

\subsection{Sum rules}

Our results are important for sum rules that relate inclusive $B\to
X_ce\bar\nu_e$ decay to the sum of exclusive channels.  For example, the
Bjorken sum rule~\cite{Bjsr} bounds the slope of the $B\to D^{(*)}e\bar\nu_e$
Isgur-Wise function, defined by the expansion
$\xi(w)=1-\rho^2\,(w-1)+\ldots\,$.  Knowing $\rho^2$ would reduce the
uncertainty in the determination of $|V_{cb}|$ from the extrapolation of the
$B\to D^{(*)}e\bar\nu_e$ decay rate to zero recoil.  This sum rule 
reads~\cite{Bjsr,IWsr}
\begin{equation}\label{bjorken}
\rho^2 = \frac14 + \sum_m\, {|\zeta^{(m)}(1)|^2 \over 4}
  + 2 \sum_p\, {|\tau^{(p)}(1)|^2 \over 3} + \ldots \,.
\end{equation}
The ellipses denote contributions from non-resonant channels.  $\zeta^{(m)}$
and $\tau^{(p)}$ are the Isgur-Wise functions for the exited
$s_l^{\pi_l}=\frac12^+$ and $\frac32^+$ states, respectively.  (For
$m=p=0$ these are the lightest orbitally excited states, and $m,p\geq1$ are
radial excitations of these.)  Since all contributions to the right-hand-side
of Eq.~\ref{bjorken} are non-negative, a lower bound on $\rho^2$ can be
obtained by keeping only the first few terms.  Using $\tau(1)\simeq0.71$ and
Eq.~\ref{ztau}, we find a large contribution to the Bjorken sum rule from the 
lightest $\frac12^+$ and $\frac32^+$ doublets,
\begin{equation}\label{bjnew}
\rho^2 > \frac14 + {|\zeta(1)|^2\over4} + 2\, {|\tau(1)|^2\over3} 
  \simeq 0.75 \,.
\end{equation}
The contribution of the $\frac12^+$ states to this bound,
which relies on the quark model result in Eq.~\ref{ztau}, is only 0.17.

The contribution of excited states to other sum rules is discussed in Ref.~[2].

\section{Conclusions}

\begin{itemize} \itemsep=-2pt  

\item At zero recoil, order $1/m_Q$ contributions to semileptonic $B\to
D_1,D_2^*$ decays (any excited charmed meson with $+$ parity) are determined by
the $m_Q\to\infty$ Isgur-Wise function and known hadron mass splittings.

\item Shape of decay spectra predicted near zero recoil, including $1/m_Q$
corrections with reasonable assumptions.

\item Large $1/m_Q$ corrections to some predictions can be tested
against experimental data in the future.

\item Better understanding of inclusive$\,=\sum\,$exclusive in semileptonic 
$B$ decay.

\item Test applicability of heavy quark symmetry for $B$ decays to
excited charmed mesons.


\end{itemize}

\section*{Acknowledgments}

I thank Mike Witherell for inviting me to a very pleasant symposium.  
The results presented here were obtained in a most enjoyable
collaboration with Adam Leibovich, Iain Stewart, and Mark Wise.  
This work was supported in part by the U.S.\ Dept.\ of
Energy under grant no.\ DE-FG03-92-ER~40701 and DOE-FG03-97ER40506 and 
by the NSF grant PHY-9457911.

\end{document}